# Bipolar resistive switching in amorphous titanium oxide thin films

Hu Young Jeong and Jeong Yong Lee

*Department of Materials Science and Engineering, KAIST, Daejeon 305-701, Korea*

Min-Ki Ryu and Sung-Yool Choi [a]

*Electronics and Telecommunications Research Institute (ETRI)*

*161 Gajeong-Dong, Yuseong-Gu, Daejeon, 305-700, Korea*

Using isothermal and temperature-dependent electrical measurements, we investigated the resistive switching mechanism of amorphous titanium oxide thin films deposited by a plasma-enhanced atomic layer deposition method between two aluminum electrodes. We found a bipolar resistive switching behavior in the high temperature region (> 140 K), and two activation energies of shallow traps, 0.055 eV and 0.126 eV in the ohmic current regime. We also proposed that the bipolar resistive switching of amorphous $TiO_2$ thin films is governed by the transition of conduction mode from a bulk-limited SCLC model (Off state) to an interface-limited Schottky emission (On state), generated by the ionic movement of oxygen vacancies.

---

[a] Electronic mail: sychoi@etri.re.kr

Recent observations of the reversible resistive switching effect in metal-oxide-metal structures (MOM) with binary oxides of $TiO_2$ [1,2] and $NiO_2$ [3,4] have been reported for next generation nonvolatile memory applications. To explain the microscopic origin of the resistive switching, various models have been proposed such as filamentary conducting paths,[5] trap-controlled space-charge-limited current (SCLC),[6] and interface Schottky contact or traps.[7] Although the versatile switching phenomena of $TiO_2$ were already reported in 1968 by Argall,[8] the resistive switching mechanism has not been clearly understood yet. Recently, we reported the resistive switching characteristics of Al/amorphous $TiO_2$ film/Al devices and the application for the resistive random access memory (RRAM).[9] In this letter, we analyze the bipolar resistive switching mechanism in Al/$TiO_2$/Al devices, by using isothermal and temperature-dependent I-V measurements.

Using a plasma-enhanced atomic layer deposition (PEALD) method, the amorphous $TiO_2$ thin films were deposited onto Al bottom electrode of 60nm thickness patterned on silicon dioxide substrates. The line-shaped Al bottom electrodes of 60 μm width was formed using metal shadow mask by thermal evaporation method on a $SiO_2$/Si substrate. To grow ultrathin $TiO_2$ films, we used titanium isopropoxide ($Ti(OCH(CH_3)_2)_4$; TTIP) and $O_2$ plasma as Ti precursor and oxygen source, respectively. We deposited amorphous titanium oxide thin films with a thickness of ~13 nm (300cycle) and a growth rate of 0.045 nm/cycle at 180 °C. The line-shaped top Al electrodes of 60 μm width were deposited perpendicularly to the bottom electrodes, so that the cross-bar type memory cell area of 60 μm ×60 μm (3600 μm$^2$) was fabricated. The I-V characteristics of the device were measured using a Keithley 4200 Semiconductor Characterization System in a DC sweep mode. The temperature-dependent electrical properties of Al/$TiO_2$/Al devices was measured from 83K to 323K. The cross-sectional transmission electron microscopy (TEM) images of Al/$TiO_2$/Al samples were obtained using a 300 kV JEOL JEM 3010 with a 0.17 nm point resolution.

Figure 1 shows a typical current density-voltage (J-V) curve of Al/TiO$_2$/Al device performed at room temperature (298K). The voltage sweep sequence was indicated by the arrows. The pronounced hysteretic behavior was clearly observed, showing the resistance ratio of > $10^2$ at a reading voltage (-1 V) without an initial forming process. Interestingly, when the positive bias was firstly applied to the device, the hysteresis did not occur, resulting in a failed state at 7V bias, as shown in the left inset of Fig. 1. Therefore, our device represents asymmetric bipolar resistance switching (BRS),[10] in which the negative bias changes the device from high-resistance state (HRS) to low-resistance state (LRS) and the positive bias from LRS to HRS. The asymmetric J-V characteristics of Al/TiO$_2$/Al devices can be explained by the asymmetric interface formation, as reported by Yu et al.[9]

The right inset of Fig. 1 shows a cross-sectional high resolution transmission electron microscopy (HRTEM) image of the fabricated layered film. The thickness of TiO$_2$ film was about ~14nm. The deposited TiO$_2$ films were amorphous phase as revealed by TEM image. Interestingly, it was also observed that the top and bottom amorphous interface layer were formed between Al metal electrodes, as indicated by bright contrast in the HRTEM image. In the case of the bottom interface, a native aluminum oxide with 2~3nm thickness was formed on the surface of Al bottom electrode. In contrast, the top interface layer was created by an out-diffusion of oxygen ions in TiO$_2$ layer, resulting in amorphous oxide layer. Thus, the inner domain of TiO$_2$ layer changed to TiO$_{2-x}$ with oxygen deficiency which causes various defects such as oxygen vacancies ($V_O^x$, $V_O^+$, $V_O^{++}$) or Ti$^{+3}$ center. These defects can act as the charge trapping sites.

In order to clarify the current conduction mechanism of Al/TiO$_2$/Al device, we analyzed the J-V characteristics with respect to various conduction mechanisms: SCLC, Richardson-Schottky (R-S) thermionic emission, and Poole-Frenkel (P-F) conduction.[11-12]

Figure 2(a) and (b) show the logarithmic plots of the J-V curve for the negative and positive voltage regions, respectively. In the first negative bias scan, shown in Fig. 2 (a), the slope of logJ-logV plot (S) of the HRS changes from 1 to 2, 20, and 2 as V increases. This behavior is in qualitative agreement with the trap-associated SCLC theory. According to SCLC model, logJ-logV plot exhibits linear ohmic conduction ($I \propto V$), followed by a square dependence on voltage: $I \propto V^2$, which corresponds to Child's law region.[13] When the applied voltage reaches to a certain threshold voltage $V_T$ (~ - 2.1 V) the current increases rapidly due to trap-filled condition. In a higher-voltage region $(V > V_T)$, $I \propto V^2$ is observed again. However, in the second negative sweep (LRS), the S values show just a single increase from 1 to 1.5. This is not simply explained by the SCLC model as generally observed in many oxide films.[6,10] Hence, it is reasonable to assume that the HRS and LRS follow two different conduction mechanisms.

The linear fittings of the $\ln(J/T^2)$ vs $E^{1/2}$ and $\ln(J/E)$ vs $E^{1/2}$ is used to determine whether the Schottky emission or the Pool-Frenkel effect is the controlling conduction mechanism in high voltage region of negative HRS and >0.1V region of negative LRS, as shown in Fig. 2 (c) and (d).[11-12] It can be seen that reasonable straight lines can be fitted in the high voltage region of LRS for both plots. The linear line slopes (S) of each plot are analyzed in Fig. 2 (c) and (d), from which we can calculate dynamic dielectric constants ($\varepsilon_r$) based on following equations, respectively

$$\varepsilon_{SC} = \frac{q^3}{(kTS)^2 4\pi\varepsilon_0}$$

and

$$\varepsilon_{PF} = \frac{q^3}{(kTS')^2 \pi\varepsilon_0}$$

Where $S$ and $S'$ is the slope of the linear fitting obtained from the $\ln(J/T^2)$ vs $E^{1/2}$ and $\ln(J/E)$ vs $E^{1/2}$, $q$ is the electronic charge, $kT$ is Boltzmann's constant times the absolute temperature, and $\varepsilon_0$ is the dielectric constant of free space. Because the extracted dielectric constant, $\varepsilon_{SC}$ ($\sim 34$), is between low-frequency (static) dielectric constant and high-frequency (optical) dielectric constant ($\varepsilon = n^2$, where n is the refractive index), Schottky emission is more likely to be the dominant conduction mechanism at the high electric field in the negative HRS and at > 0.1V region in the negative LRS. It is remarkable that our data is closer to the static dielectric value of $TiO_2$ ($\varepsilon \sim 40$) [14] which is mainly contributed by ionic polarizability. Thus, it is believed that the change of current conduction mode of a $Al/TiO_2/Al$ stack is associated with the ionic movement like oxygen vacancies (ions). Although the transition from a bulk-limited conduction of SCLC to an electrode-limited conduction of Schottky typically occurs in high voltage regions,[12,15] the reversible change of two conduction modes, accompanying the resistance change, has not yet reported.

To investigate the details of the defect (oxygen vacancy) characteristics in the amorphous titanium oxide films, temperature dependent I-V curves were measured in the temperature range from 83 to 323 K (Figure 3(a)). One notable finding is that the larger hysteresis was observed as temperature increases and a negligible switching behavior was shown in the low temperatures (T<140K). It is believed that temperature makes a critical effect in the resistive switching of amorphous thin films. This phenomenon can be understood with the assumption that the diffusivity of oxygen vacancies (ions) is also dependent on the temperature. In the low temperature, the movement of oxygen ions in the oxide lattice would not be easy. In order to deeply estimate the activation energies ($E_a$) of traps, the temperature dependence of the conductivity ($\sigma = \sigma_0 \exp(-E_a/kT)$) at the ohmic current regions (- 0.1V ~ + 0.1V) was plotted in Fig. 3(b). The activation energies were obtained from the slope of the least-squared fit lines. In the low temperature regions (T < 140 K), the value of $\Delta E$ of both resistance states is the same as 0.027 eV. However, in the high temperature (T > 140 K), the activation energies of both resistance states are separated. The energy level of high resistance state is 0.055 eV and that of low resistance state is 0.126 eV. These values are the activation energies of the shallow traps located near the conduction band edge. It is remarkable that the different resistance states also have unlike activation energies. However, it seems doubtful that low resistance state has a high activation energy. This is probably associated with the dissimilarity of the defect (oxygen vacancy) concentrations present on the inner bulk $TiO_{2-x}$ layer in two different resistance states. Cronemeyer[16] reported that the electrical conductivity emerged from the ionization of either one or two trapped electrons from each oxygen vacancy and the decrease of thermal ionization appeared with an increase of oxygen vacancy concentration. The change of the activation energy of these shallow traps might be originated from that of the trap (oxygen vacancy) density in the bulk $TiO_{2-x}$ caused by oxygen vacancy drift. On the basis of this result, we can explain that the low resistance state with traps of high activation energy could be exist due to low oxygen vacancy concentration.

Based on above results, the schematic band diagrams of bipolar resistive switching in the amorphous titanium oxide thin films was illustrated in Fig. 4. During the top Al electrode deposition on the amorphous titanium oxide thin films, the top interface layer is formed by the redox reaction between Al and $TiO_2$ film, resulting in oxygen-rich region at the interface and oxygen-deficient region in the inner $TiO_2$ layer, respectively, as shown in Fig. 4(a). The interface region with a large amount of oxygen ions makes accumulation-type contact with Al metal electrode due to lower Fermi level ($E_{TiO_2}$) compared to Al work function. However, when the high negative bias applies to the top electrode, the oxygen vacancies present on the inner $TiO_{2-X}$ domain can drift into the top interface. During this process, the interface region accumulated with oxygen vacancies changes into highly doping region, therefore Schottky contact mode can be induced.

In Summary, we successfully fabricated the Al/amorphous-$TiO_2$/Al resistance switching devices by using a plasma-enhanced atomic layer deposition method. We observed the hysteretic and asymmetric I-V characteristics in our symmetric devices, caused by the asymmetric interface formation process. Detailed analyses of isothermal I-V characteristics revealed that the bipolar resistance behavior is dominated by the transition of the conduction mechanism in the vicinity of Al/$TiO_2$ top interface, from the bulk trap-controlled SCLC to interface-controlled Schottky emission, accompanied by oxygen vacancy(or ion) diffusion. From the temperature-dependent I-V measurements, we found two separated trap activation energies (0.055 eV and 0.126 eV) in the high temperature (T > 140K) region. We suggested that difference of electron activation energy originated from that of the concentration of oxygen vacancies in the inner top region.

**Acknowledgments**

This work was supported by the Next-generation Non-volatile Memory Program of the Ministry of Knowledge Economy, Korea. The authors sincerely thank the technical staffs of the Process Development Team at ETRI for their support with the PEALD facility.

**Figure Captions**

FIG. 1. (Color online) Typical J-V curve of our Al/TiO$_2$/Al device. The arrows in the figure depict the direction of the voltage sweep. The left inset shows I-V curves when the positive sweeps firstly applied to Al top electrode. The right inset shows the cross-sectional HRTEM image of Al/TiO$_2$/Al sandwiched structure.

FIG. 2. (Color online) A double-logarithmic plot of Al/TiO$_2$/Al devices in (a) negative bias and (b) positive bias regions. (c) The Schottky plot and (d) Poole-Frenkel plot in a negative bias region. The S values mean the slopes of linear fitting region. $\varepsilon_{SC}$ and $\varepsilon_{PF}$ denote the dielectric constants extracted from Schottky emission and Poole-Frenkel conduction models.

FIG. 3. (Color online) (a) The temperature-dependent I-V characteristic (from 83K to 323K) of a Al/TiO$_2$/Al structure. (b) The temperature dependences of conductivity obtained in the ohmic current region (-0.1V~+0.1V) with the two different states (On/Off).

FIG. 4. (Color online) (a) Accumulation contact formed at Al/TiO$_{2-x}$ interface by redox reaction between Al and TiO$_2$, corresponding to HRS. (b) Schottky contact induced by the diffusion of oxygen vacancies into interface region, corresponding to LRS.

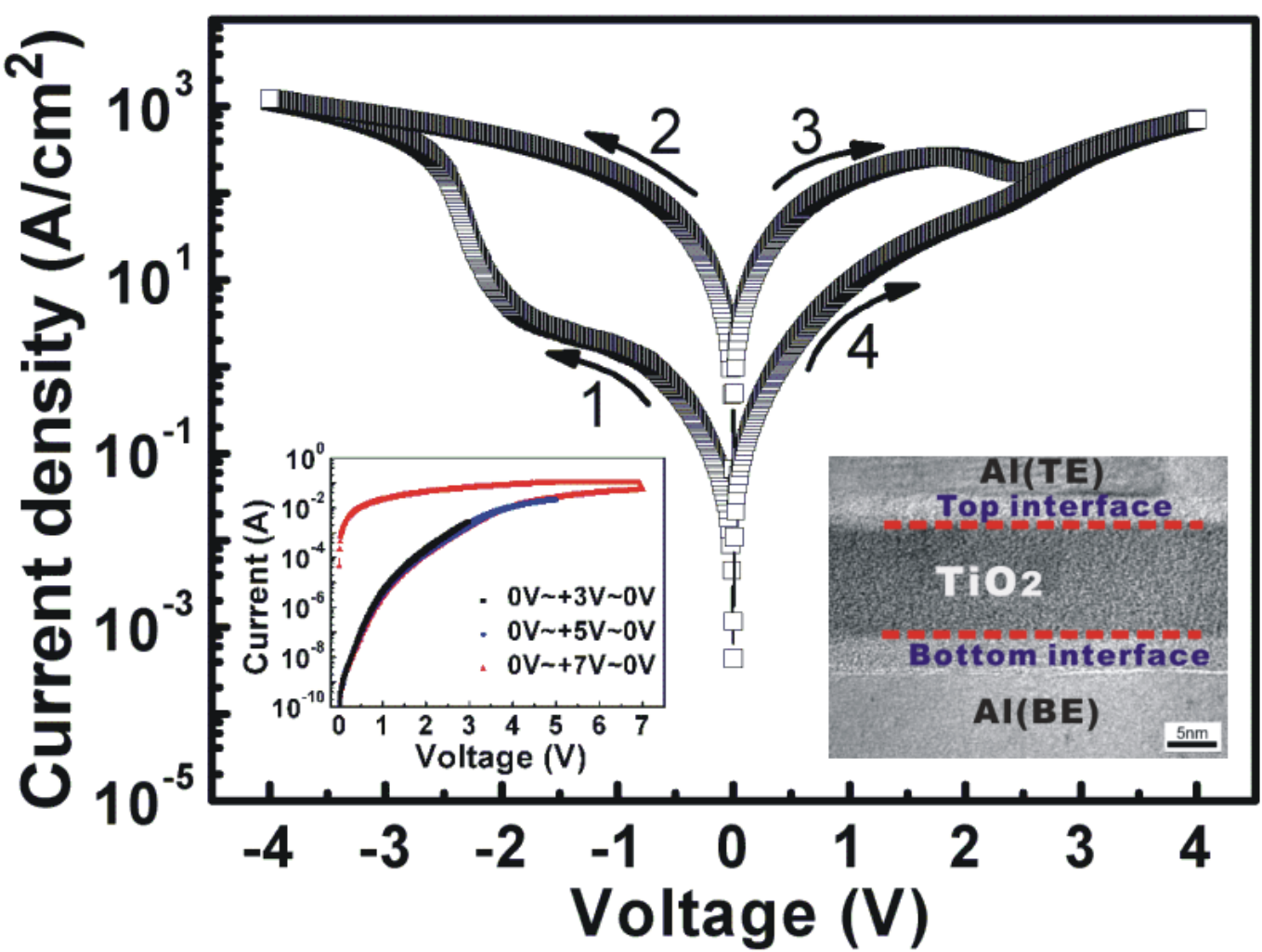

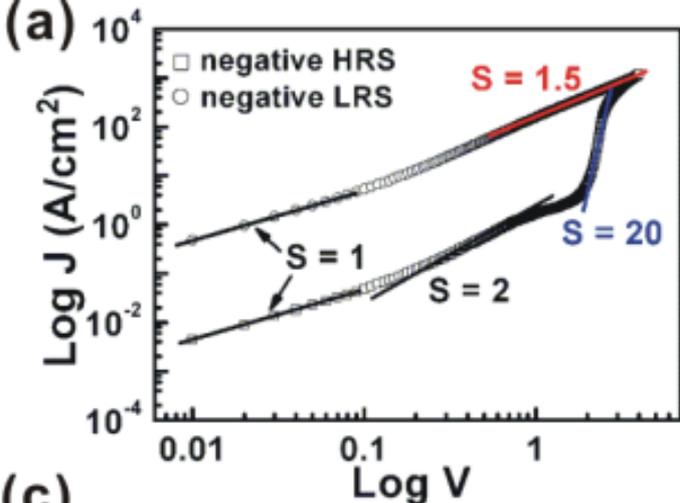
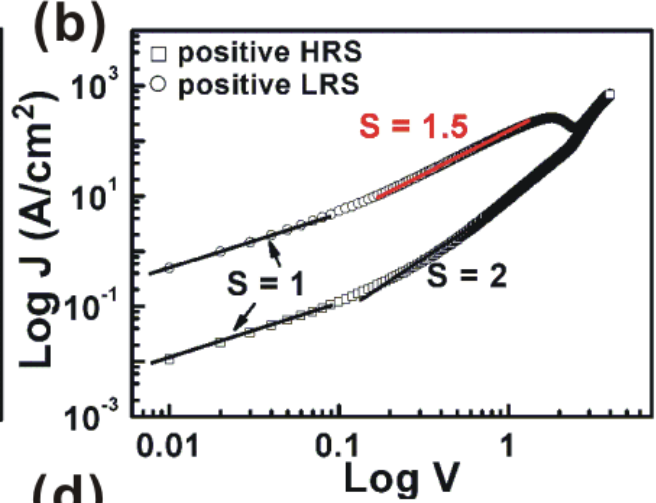
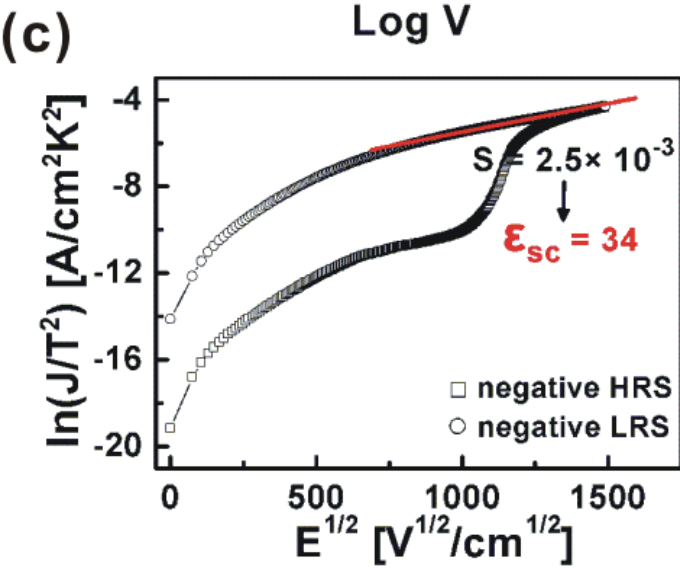
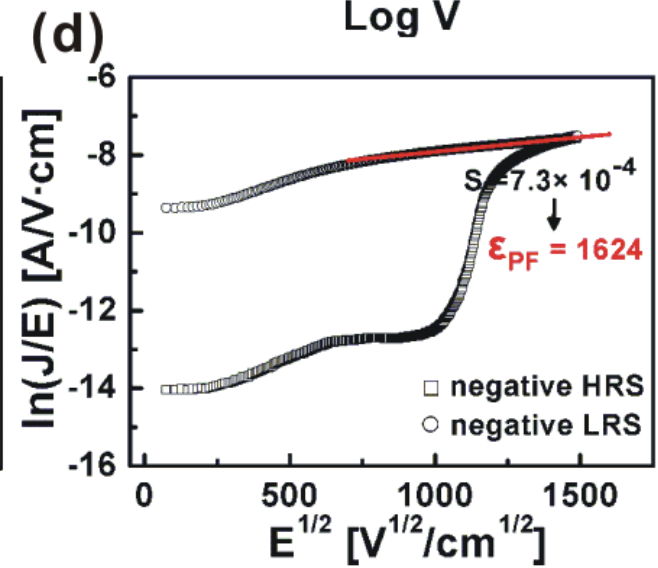

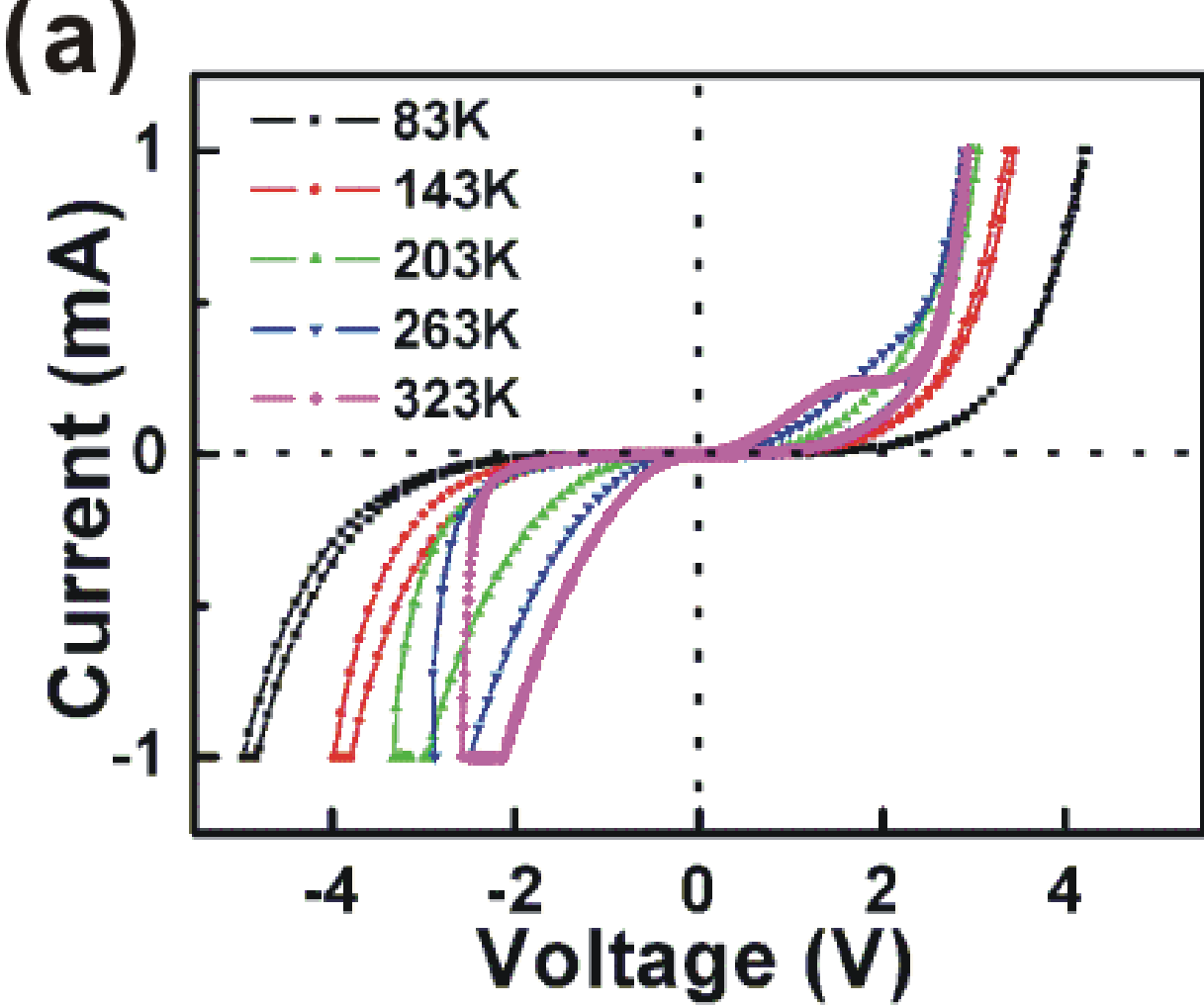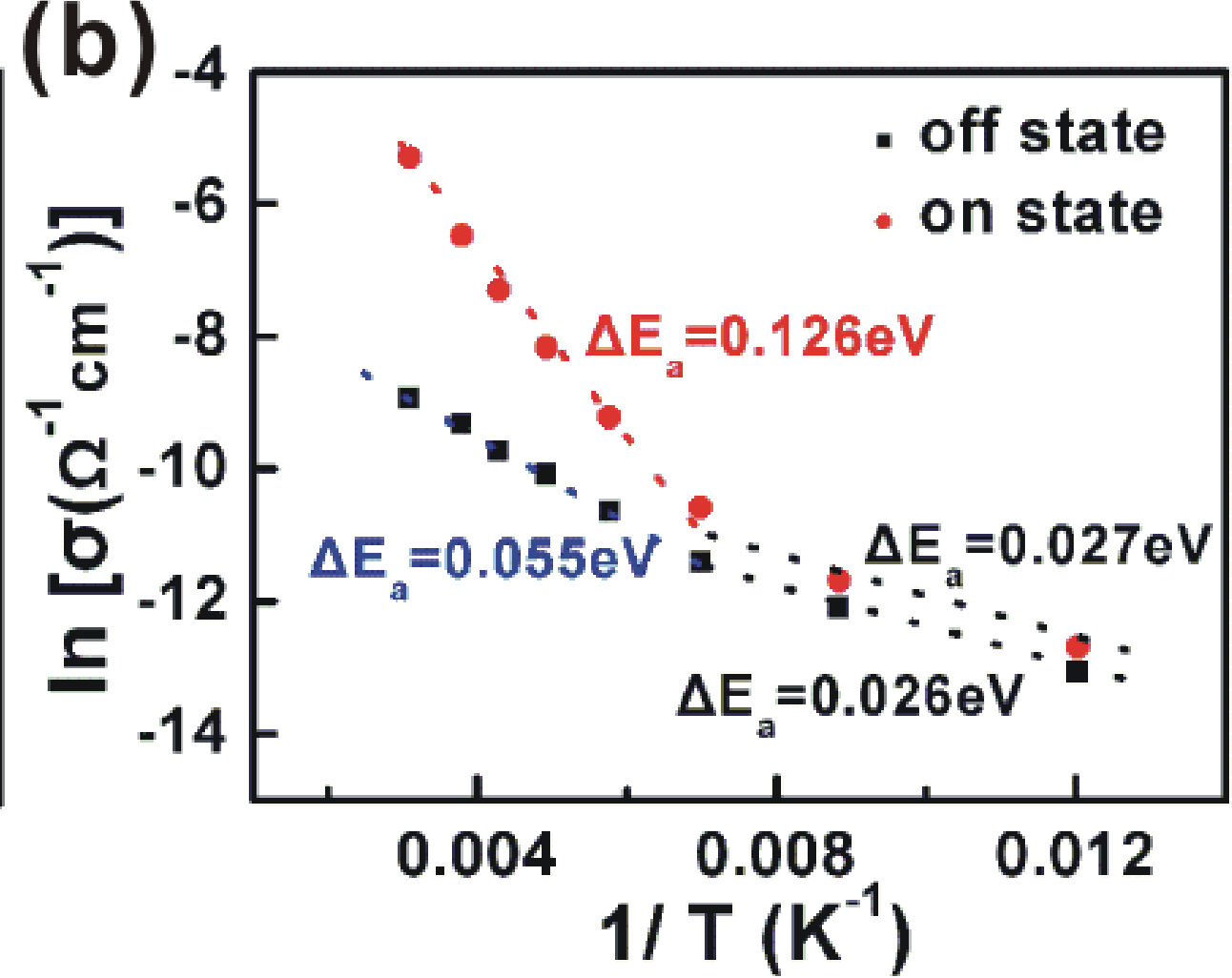

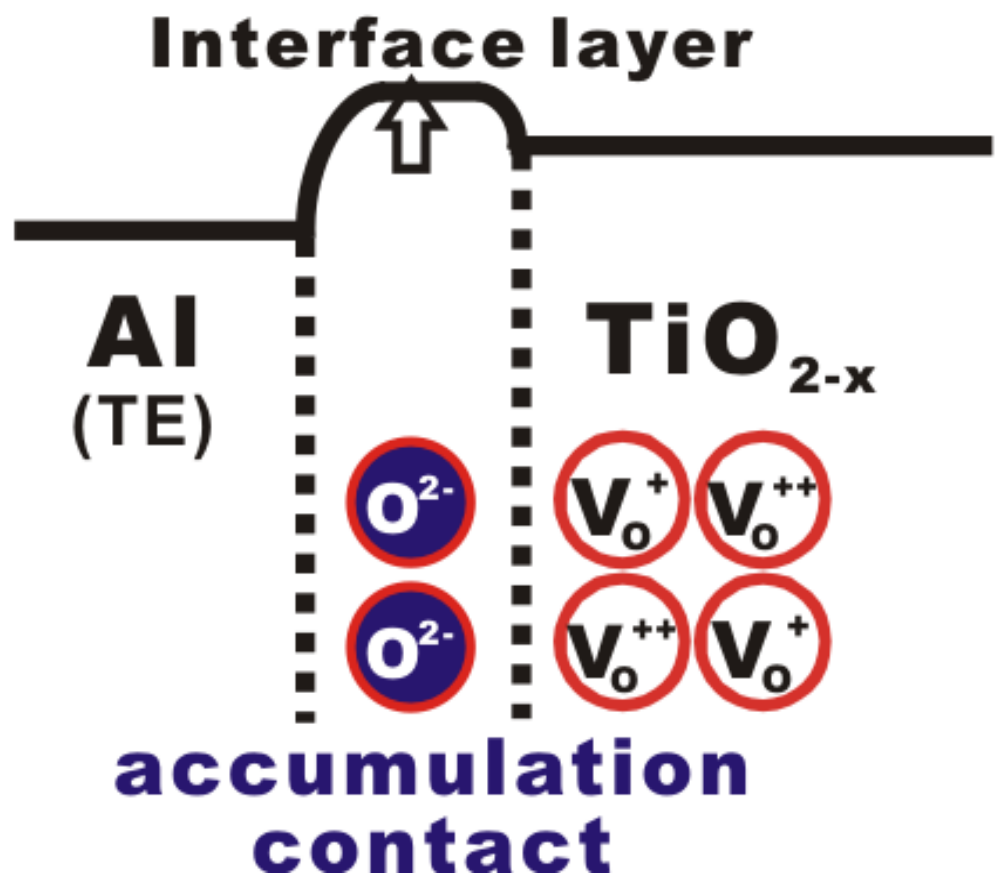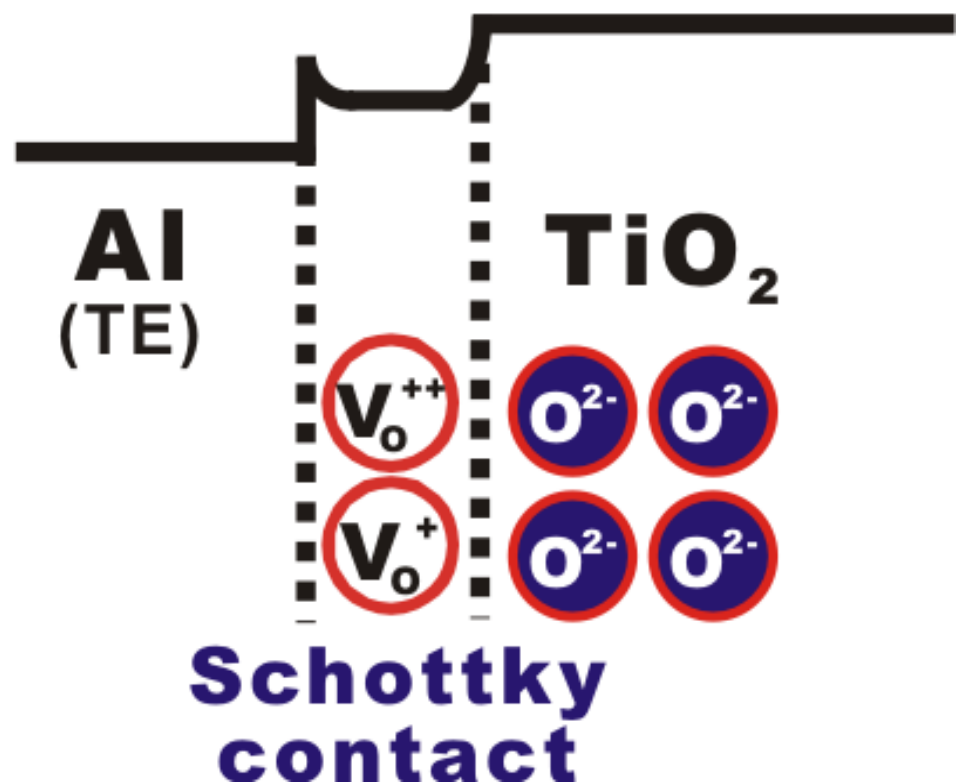